\documentclass[useAMS,usenatbib]{mn2e}
\usepackage{times}
\usepackage{color}
\usepackage{amsfonts,amssymb,amsmath}
\usepackage{aas_macros}
\usepackage{graphicx}
\title[EA-based analysis of microlensing lightcurves]{Evolutionary algorithm-based analysis of gravitational microlensing lightcurves}
\author[V.~Rajpaul]{V.~Rajpaul$^{1,2}$\thanks{E-mail: vinesh.rajpaul@uct.ac.za}\\
$^{1}$Astrophysics, Cosmology and Gravity Centre (ACGC), University of Cape Town, Private Bag X3, Rondebosch 7701, South Africa\\
$^{2}$South African Astronomical Observatory, PO Box 9, Observatory, 7935, South Africa}
\begin{document}
\date{Accepted 2012 August 16.  Received 2012 August 15; in original form 2012 February 23}
\pagerange{\pageref{firstpage}--\pageref{lastpage}} \pubyear{2012}
\maketitle
\label{firstpage}
\begin{abstract}
A new algorithm developed to perform autonomous fitting of gravitational microlensing lightcurves is presented. The new algorithm is conceptually simple, versatile and robust, and parallelises trivially; it combines features of extant evolutionary algorithms with some novel ones, and fares well on the problem of fitting binary-lens microlensing lightcurves, as well as on a number of other difficult optimisation problems.  Success rates in excess of $90\%$ are achieved when fitting synthetic though noisy binary-lens lightcurves, allowing no more than 20 minutes per fit on a desktop computer; this success rate is shown to compare very favourably with that of both a conventional (iterated simplex) algorithm, and a more state-of-the-art, artificial neural network-based approach. As such, this work provides proof of concept for the use of an evolutionary algorithm as the basis for real-time, autonomous modelling of microlensing events. Further work is required to investigate how the algorithm will fare when faced with more complex and realistic microlensing modelling problems; it is, however, argued here that the use of parallel computing platforms, such as inexpensive graphics processing units, should allow fitting times to be constrained to under an hour, even when dealing with complicated microlensing models. In any event, it is hoped that this work might stimulate some interest in evolutionary algorithms, and that the algorithm described here might prove useful for solving microlensing and/or more general model-fitting problems. \end{abstract}
\begin{keywords}
gravitational lensing: micro -- binaries: general -- methods: numerical
\end{keywords}
\section{Introduction}
Gravitational microlensing is an established technique for detecting exoplanets. When a massive foreground object (the lens, e.g.\ a planet and its host star) passes in front of a distant, background star (the source), the latter is magnified and displays a characteristic microlensing lightcurve. Since the lens is detected on account of its mass rather than its luminosity, faint planetary-mass objects can be detected by microlensing. Indeed, microlensing has the potential to yield the most representative statistical sample of Milky Way planets -- unlike many complementary techniques used to detect exoplanets, it is in principle sensitive enough to detect even very distant, Earth-mass planetary objects \citep{Bennett:1996,Wambsganss:2011}. Unfortunately, microlensing events are extremely rare, requiring a very precise alignment between observer, lens and source: as of July 2012, of the several hundred known exoplanets, only around a dozen were discovered by microlensing (see \citealp{Shvartzvald:2012}, and references contained therein). Still, many of these detections have constituted important discoveries in the broader context of exoplanetary science \citep[e.g.][]{Beaulieu:2006,Gaudi:2008,Cassan:2012} -- \textcolor{black}{recently, it has even been suggested that microlensing has facilitated the detection of a number of free-floating, planetary mass objects in the Galaxy (\citealt{Sumi:2011}; cf., however, \citealt{Quanz:2012}).}

Very comprehensive models do exist for describing microlensing events and their corresponding lightcurves, though unfortunately it is notoriously difficult to use these models to interpret microlensing events: amongst other complicating factors, the models tend to be highly nonlinear, and have enormous parameter spaces that are often fraught with ambiguities and degeneracies \citep{Dominik:1999,Vermaak:2007}. Even the simplest possible microlensing model (viz.\ a point-like source star lensing an isolated mass) poses some nontrivial challenges to microlensing modellers \citep{Dominik:2008,Dominik:2009}.

This paper presents a new metaheuristics algorithm which combines features of extant evolutionary algorithms (genetic algorithms; evolution strategies) with some novel ones, developed with a view to performing efficient and autonomous fitting of (especially binary-lens) microlensing lightcurves.  The algorithm is, however, robust enough to solve general nonlinear optimisation problems, and its development was informed by tests carried out on a broad class of optimisation problems.

Section 2 of the paper gives an overview of the binary-lens model used to test the performance of the new algorithm; Section 3 describes evolutionary algorithms in general, as well as the new algorithm which is the focal point of this work (Appendix A, at the very end of the paper, provides a more detailed look at the `nuts and bolts' of the new algorithm); Section 4 focuses on the fitting experiments (and their results) used to assess the new algorithm; and Section 5 contains a commentary on the results presented in the preceding section. Section 6 concludes.
\section[Overview of binary-lens model]{Overview of binary-lens model}
\begin{figure}
\begin{center}
\includegraphics[scale=0.85]{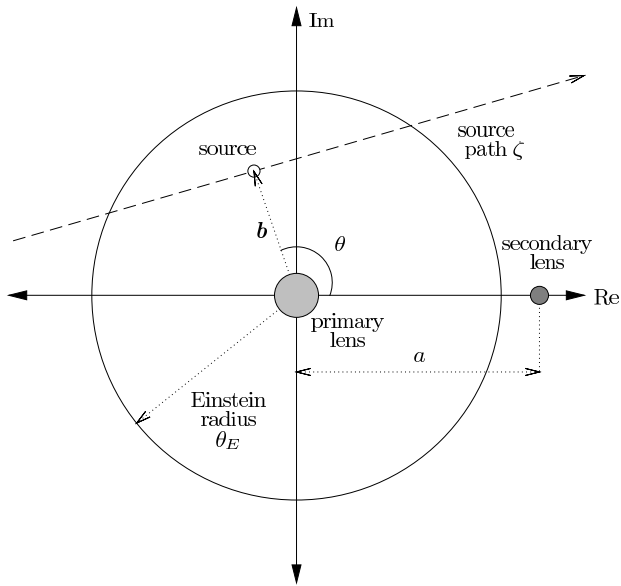}
\caption{Geometry assumed in the SBLM.}
\label{fig:lensGeom}
\end{center}
\end{figure}
A small companion to a stellar lens can be detected via perturbations it introduces to the lightcurve expected for an isolated lens. The resulting binary-lens lightcurve can exhibit a very wide variety of morphologies: it might be practically indistinguishable from a point-lens lightcurve, or it might exhibit complex structure including significant asymmetry, multiple peaks, and spikes of high (formally infinite) magnification produced during so-called `caustic crossings' \citep*{Mao:1991,Night:2008}. Models of such binary-lens events are particularly useful for characterising exoplanetary microlensing events: even a system with one star and multiple planetary bodies can often be well-approximated either by ignoring multiple planets, or by treating each planet plus its host star as an independent binary system, provided source magnification is not too high \citep*{Gaudi:1998}.

The model introduced here features $7$ basic parameters that describe rectilinear motion of an unblended, point-like source across a static binary lens. This simple binary-lens model (hereafter SBLM) neglects some higher-order effects that need to be taken into account when carrying out in-depth modelling of binary-lens events: nevertheless, the model is far from trivial, and is useful enough to provide first-order fits to many binary-lens events. 

For example, even though the assumption of a point-like source will break down in a large fraction of real events \citep{Dominik:2008}, the regions of lightcurves affected by finite-source effects tend to be localised -- usually affecting the high-magnification peaks associated with e.g.\ caustic crossings -- so even when such effects are present, good SBLM-based fits can often be obtained simply by excluding the relevant regions of lightcurves from the fitting \citep{Vermaak:2007}. It should be emphasised, however, that this work does not attempt to advocate the SBLM for use in modelling real microlensing events: the model is adopted here primarily to provide a relatively straightforward platform for benchmarking different algorithms studied later in this work (see Section \ref{section:experiments}).
\subsection{Model parameters}
A useful scale for characterising a lensing event is the so-called Einstein radius or Einstein angle of the primary lens, $\theta_E$, defined as:
\begin{equation}
{\theta _E} := \sqrt {\frac{{4GM}}
{{{c^2}}}\left( {\frac{{{D_S} - {D_L}}}
{{{D_L}{D_S}}}} \right)}, 
\end{equation}
where $G$ is the gravitational constant, $M$ is the mass of the primary lens, $c$ is the vacuum speed of light, $D_L$ is the distance between observer and the lens, and $D_S$ is the distance between observer and the source. The Einstein radius is the angular radius of the ring that would be formed in the case of perfect source-lens-observer alignment; for typical Bulge microlensing events, $\theta_E\sim1~\textrm{mas}$.

For convenience, complex notation is used here to describe the lensing event. The origin of the coordinate axes is placed at the projected position of the primary lens; with no loss of generality the secondary lens is placed on the real ($+x$) axis; and $\zeta\in\mathbb{C}$ is used to denote the position of the source on the sky.  The SBLM's seven parameters, and their physically-permissible ranges, then, are as follows (refer also to Fig.\ \ref{fig:lensGeom}):
\begin{enumerate}
	\item $a$, the projected angular orbital separation, in units of $\theta_E$, between the primary and secondary lenses ($0<a<\infty$);	 
			\item $b$, the length, in units of $\theta_E$, of the projected source-lens impact vector $\bmath{b}$ ($0<b<\infty$);
	\item $m_0$, the unlensed magnitude of the source;
	\item $q$, the ratio of the mass of the secondary lens to that of the primary lens ($0<q<1$);
	\item $\theta$, the angle formed between the $+x$ axis and the projected impact vector $\bmath{b}$ ($0<\theta<2\pi$);
	\item $t_E$, the Einstein radius crossing time, i.e.\ the \textcolor{black}{time it takes} the source to travel an angular distance of $\theta_E$; and
	\item $t_m$, the time of closest projected approach of the source to the primary lens.
\end{enumerate}
Note that to obtain physical angular distances, the parameters $a$ and $b$ must be scaled by $\theta_E$.

Many other parametrizations of binary lensing events are possible \citep[see e.g.][]{Mao:1995,Dominik:1999}, although most are qualitatively similar to the one used here. In all subsequent discussions, the seven SBLM parameters will be assumed to be constrained to the ranges given in Table \ref{tab:params}. These ranges were chosen to facilitate direct comparison with the results of \citet{Vermaak:2003,Vermaak:2007}; they cover most of the lensing zone, and thus the geometries with the highest likelihood of detecting secondary lenses.

\begin{table}
\centering
\caption{Allowed parameter ranges for all fits and simulated lightcurves.}
\label{tab:params}
\begin{tabular}{cccc}
\hline
Parameter & Units &    Minimum &    Maximum \\
\hline
       $a$ & $-$ &      $0.6$ &      $1.7$ \\

       $b$ & $-$ &    $0.001$ &        $1$ \\

     $m_0$ & mag &       $18$ &       $22$ \\

       $q$ & -- &      $0.1$ &        $1$ \\

  $\theta$ & rad &        $0$ &      $2\pi$ \\

     $t_E$ & d &        $5$ &       $50$ \\

     $t_m$ & d &      $-20$ &       $20$ \\
\hline
\end{tabular}  
\end{table}

\subsection{Calculation of binary-lens lightcurves}
One of the primary difficulties associated with the SBLM is that it does not provide an explicit expression for the amplification of the source as a function of time and lensing parameters. Instead, the time-varying source position relative to the lens, as well as the lens geometry, is first used to calculate the position of the \emph{images} of the source that are formed by the lens; the light from each of these images is then summed to calculate the total amplification.

Assuming rectilinear motion, straightforward geometric considerations yield the \emph{source} position as a function of time, and two of the lensing parameters:
\begin{equation}\label{eq:sourcePos}
\zeta  = \tau\sin \theta  + b\cos \theta  + i\left( { - \tau\cos \theta  + b\sin \theta } \right),
\end{equation}
where $\tau$ is a dimensionless time parameter:
\begin{equation}
\tau := \frac{{t - {t_m}}}{{{t_E}}}.
\end{equation}
Assuming zero external gravitational shear and a binary lens, the general lens equation \citep*[see][]{Bourassa:1973,Witt:1990} reduces to
\begin{equation}\label{eq:lensing}
\zeta  = z - \frac{1}{{\bar z}} + \frac{q}{{a - \bar z}},
\end{equation}
which, with $\zeta$ known, is an implicit expression for the \emph{image} positions, $z\in\mathbb{C}$. It may be shown that this special case of the lensing equation has always either $n=3$ or $n=5$ solutions, corresponding to 3 or 5 lensed images. Following some nontrivial algebraic manipulation \citep{Witt:1990}, Eqn.\ \ref{eq:lensing} can be converted into a complex quintic polynomial equation of the form
\begin{equation}\label{eq:lensingPoly}
f(z;\zeta ,a,q) = \sum\nolimits_{j = 0}^5 {{c_j}(\zeta ,a,q)  {z^j} = 0},
\end{equation}
where $c_j\in\mathbb{C}$; the roots of this polynomial can be found using any one of a number of standard numerical techniques, for example the Jenkins-Traub algorithm or Laguerre's algorithm (in practice it is more convenient to solve this polynomial equation than trying directly to find all the solutions to Eqn.\ \ref{eq:lensing}). Two of the five polynomial roots may not be true solutions to the lensing equation itself, but these spurious solutions may readily be eliminated by direct substitution into Eqn.\ \ref{eq:lensing}.  In an alternative though equivalent formulation, the lensing equation can be cast into the form of a real polynomial equation \citep{Asada:2002}.

Once the image positions $z_i$ have been solved for, the amplification due to the $i^{\rmn{th}}$ lensed image is calculated via
\begin{equation}
{A_i} = \frac{1}{{\det \left| {{\mathsf{J}_i}} \right|}} = {\Biggl| {1 - \biggl| {\frac{1}{{{z_i}^2}} + \frac{q}{{{{(a - {z_i})}^2}}}} \biggr|} \Biggr|^{ - 1}},
\end{equation}
where $\mathsf{J}_i$ is the Jacobian of Eqn.\ \ref{eq:lensing} for the case of the $i^{\rmn{th}}$ image. Finally the amplification due to the $n$ individual images is summed, so that the lensed magnitude $m$ is given by:
\begin{equation}\label{eq:A_s}
m^{(k)} =  - 2.5\cdot{\log _{10}}\left( {\sum\nolimits_{i = 1}^n {{A_i^{(k)}}} } \right) + {m_0};
\end{equation}
the superscript $k=1,2,\ldots,N$ is introduced to emphasise that the preceding calculations culminate in a single point on an $N$-point lightcurve.

\subsection{Difficulties associated with fitting binary lightcurves}\label{sec:difficulties}
Given a set of parameters, obtaining the lightcurve predicted by the SBLM is a nontrivial though relatively straightforward task. Unfortunately, however, the associated \emph{inverse problem} -- i.e., given a lightcurve, using the SBLM to determine the physical parameters associated with the lensing event -- is extremely difficult; yet it is of course the solution to this inverse problem that is of value insofar as the interpretation of real-world microlensing data is concerned. Some of the difficulties associated with this fitting problem include the following.
\begin{enumerate}
	\item \emph{Computational complexity}. Generating even a single $N$-point lightcurve, i.e.\ solving the \emph{forward problem}, requires, as a minimum, the roots of $N$ different quintic polynomials to be found numerically \citep[for techniques on solving quintic polynomial equations, see e.g.][]{King:1996,Ralston:2001,Press:2007}.
	\item \emph{Volume of parameter space}. The SBLM has a 7-dimensional parameter space, and the physically-realistic ranges for some of the parameters (e.g.\ impact parameter, lens mass ratio) can span many orders of magnitude.  \citet{Vermaak:2003} estimated that an exhaustive grid search of the parameter space would take on the order several of years to complete (assuming a $2\%$ error tolerance on all parameters, the very conservative parameter ranges in Table \ref{tab:params}, and $\sim2$~ms/lightcurve calculation). Any extensions to the SBLM only exacerbate this problem.
	\item \emph{Nonlinearity}. The mapping from SBLM parameter space to source amplification is highly nonlinear -- consequently, nearly-identical parameter sets can give rise to dramatically different lightcurves. A typical regression surface will contain a large number of local optima, will be non-smooth, and will contain no clue as to where the global optima (with respect to some, e.g.\ $\chi^2$-based, metric) are to be found; furthermore the wells of convergence around optima tend to be small \citep{Vermaak:2003,Bennett:2010}, and when dealing with noisy data, true solutions need not even correspond to a globally optimal solutions. Using biased parameter estimators (e.g.\ using a maximum likelihood estimator, rather than, say, a posterior-mode estimator) can also lead to globally optimal solutions that are, potentially, nowhere near the true solution  (\citealp{Dominik:2008}; see also Section \ref{section:experiments} of this paper).
	\item \emph{Degeneracy}. The aforesaid problem of similar parameters leading to very different lightcurves can be overcome by making a denser sampling of the parameter space; more challenging, however, is the problem that a number of very \emph{different} parameter sets can give rise to virtually indistinguishable lightcurves \citep{Dominik:1999}. In practice, such degeneracies can either be resolved by using non-photometric (e.g.\ spectroscopic) data to reduce the effective parameter space, or they can be avoided with high quality photometry and dense lightcurve sampling \citep{Mao:1995,Bennett:2008}.\footnote{Some of the degeneracies identified by \citet{Dominik:1999} will not even be resolved with any currently-achievable data qualities and sampling rates \citep{Shvartzvald:2012}.}
	\end{enumerate}
The impasse, then, is that there is little hope of finding the `correct' set of parameters to describe a microlensing event if one does not make a dense and extensive search of the parameter space, but the computational burden of doing so is often crippling. A thorough  search is especially important because even if one does happen to find one set of parameters that seems to provide a very good description of the event in question, there may exist many other parameter sets that provide equally good or better descriptions of the event. A good search algorithm should, therefore, be able to locate, in a reasonable amount of time, all \emph{potential} solutions; actually choosing between competing model parameter sets is a different problem altogether \citep{Burnham:2002}.
\begin{figure*}
\begin{center}
\includegraphics[scale=0.75]{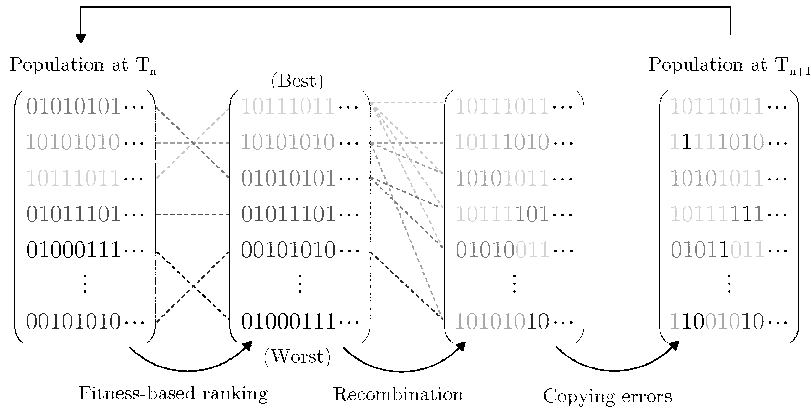}
\caption{Schematic to illustrate the workings of the canonical genetic algorithm, where trial solutions are encoded as binary strings. Each bit represents a gene; here, lighter shades are used to represent solutions determined to be fitter, according to any problem-specific metric. The `initial' population (first array) contains a range of solutions, some good (lighter shades) and some bad (darker shades). During ranking, the solutions are sorted from best to worst (second array). The genes from fitter solutions are then favoured during recombination, which is apparent from the fact that the new solutions (third array) are assembled mainly from components of the original lighter-coloured solutions; finally, random mutations -- which may or may not be beneficial -- are introduced (fourth array), and the whole process repeats.  The fine-grained details of many EAs differ from the simple algorithm illustrated here; however, most EAs share the same underlying concept of a population of trial solutions evolving, under the action of a few simple evolutionary operators, towards optimality (the constitution of the or an `optimal' solution will depend on the problem at hand).}
\label{fig:EA}
\end{center}
\end{figure*}
\subsection{A simple noise model}
A simple model of photometric noise was developed, during the course of this work, based on fits to $1000$ microlensing events observed during the 2011 campaign of the Optical Gravitational Lensing Experiment \citep{Udalski:2003}; the full dataset comprised approximately \hbox{$1.5$~million} \hbox{$I$-band magnitudes}, along with estimated photometric errors, seeing estimations and sky levels.\footnote{The data are available online at http://ogle.astrouw.edu.pl.} The best-fitting model for photometric errors as a function of $I$-band magnitude (based on data in the range \hbox{$14\la m_I^{(k)}\la22$}) thus obtained was
\begin{equation}
\label{eq:noiseModel}
{\log _{10}}{\sigma _I^{(k)}} = 0.3416 \cdot {m_I^{(k)}} - 7.7095,
\end{equation}
where $\sigma _I^{(k)}$ is to be interpreted as the standard deviation of a Gaussian distribution of observed magnitudes around some unobserved true magnitude $m_I^{(k)}$.

This model seems overly optimistic for bright sources (i.e.\ $m_I^{(k)}\sim14$, where the model predicts photometric errors on the order of $0.1\%$!), and even the assumption of Gaussian noise is probably na\"ive \citep{Dominik:2008}; still, the model does at least provide a means of making the simulated lightcurves used for fitting experiments (Section \ref{section:experiments}) somewhat more realistic and more challenging to fit. 
\section[Evolutionary algorithms]{Evolutionary algorithms}
\subsection{Overview}
Evolutionary algorithms (hereafter EAs), are metaheuristic optimisation algorithms which tend to yield good results on a wide range of (even extremely difficult) mathematical optimisation problems. EAs draw inspiration from evolutionary biology -- especially population genetics -- and they incorporate, in a computational setting, notions such as natural selection/survival of the fittest, genetic recombination, inheritance, and mutation. The first EA-based mathematical optimiser was proposed in the mid-1970s \citep{Holland:1975}, and since then, many modifications and improvements to the basic algorithm have been developed, including mechanisms without any direct biological analogues \citep{Haupt:2004}.

So-called \emph{genetic algorithms} (GAs) form one of the most successful subsets, and certainly the best-known subset, of evolutionary algorithms; \emph{evolution strategies} (ESs), developed independently from (though more or less concurrently with) GAs, form another well-known subset. In spite of the rich variety of their potential incarnations, most EAs share a basic working scheme.

As an example, consider a typical GA. The algorithm starts with a large, randomly-generated population of candidate solutions (called individuals or phenotypes) to the optimisation problem at hand, and associates with each solution an encoded version of the phenotype (called a chromosome, genotype or an individual's genetic material), as well as a problem-specific measure of the solution's quality (fitness). Then, by repeated application of `genetic operators' mainly at the genotypic level, the algorithm causes the population as a whole to increase in phenotypic fitness -- that is, solutions evolve towards optimality.  

A typical (though simplistic and by no means general or optimal) working scheme for a GA is as follows:
\begin{enumerate}
	\item construct a random initial population of genotypes;
	\item decode the genotypes and evaluate their phenotypic fitness; if the fittest phenotype matches the user-defined target fitness (or other termination criterion), terminate, otherwise continue;
	\item produce offspring by stochastic selection and recombination of genetic material in the current population, favouring the genes of individuals with high phenotypic fitness;
	\item introduce, with some low probability, random changes (copying errors) into the genetic material of the offspring;
	\item replace low-fitness members of the old population with the offspring created in the previous step, and return to step (ii).
\end{enumerate}
This scheme is illustrated in Fig.\ \ref{fig:EA}. The selective recombination of the population's genetic material exploits information associated with good solutions to build even better ones, and the random mutations serve to inject entirely new and potentially favourable material into the gene pool that could not be obtained simply by recombining the genetic material of existing individuals. Such an evolutionary scheme has a number of features which distinguish it from random heuristics, albeit that it might bear superficial resemblance to e.g.\ a standard Monte Carlo approach \citep{Gregory:2005}. 

It may be shown that, subject to a few reasonable assumptions, the canonical GA will always converge to the global optima in the search space in question \citep*{Eiben:1990,Michalewicz:1996}; moreover, it is also possible, though usually not straightforward, to estimate the \emph{rates} at which solutions are likely to be located on given problems \citep[e.g.][]{Holland:1975,Thierens:1994}.

ESs are similar to GAs in many respects, although usually they avoid solution encoding or discretisation, evolve the algorithm's control parameters in tandem with the trial solutions, and/or use more sophisticated mutation schemes \citep{Kramer:2010}.

It should be noted that in many problems -- microlensing modelling included -- one usually cannot hope to arrive at a unique or clear-cut, globally-optimal solution, and one must instead try to sample all possible optima in a multimodal parameter space. Even if an EA is not specifically set up to search for multiple solutions, mutation operators ensure exploration of the entire parameter space (provided the evolutionary sequence is sufficiently long) -- as such, given a completed evolutionary sequence, multiple possible solutions can be obtained simply by extracting all explored phenotypes that meet some fitness criterion (as opposed to extracting only the fittest phenotype in the final population state, which should correspond to a global optimum). Moreover, if one knows beforehand that one is dealing with multimodal search spaces, simple modifications can be made to make EAs more efficient at searching for multiple solutions. A simple approach is to begin a new evolutionary sequence whenever the population starts to stagnate around an optimum; alternatively, multiple independent populations can be evolved concurrently. A more sophisticated approach might be to dedicate a fraction of the evolutionary population(s) to global exploration of the search space, and the remaining fraction purely to the exploitation of promising solutions already discovered \citep{Tsutsui:1997}.
\subsection{Advantages of evolutionary algorithms}
EAs in general offer a number of benefits that suggest their suitability for dealing with the challenges discussed in Section \ref{sec:difficulties}; some of these are outlined below.
\begin{enumerate}
\item \emph{Robustness}. EA-based optimisers can handle problems where the spaces to be searched for optima are multimodal, have low-contrast or have a very high dimensionality \citep{Charbonneau:1995}.

\item \emph{Simplicity}. In order to solve a given optimisation problem, most `off-the-shelf' EAs require only a single, unambiguous measure of the quality (fitness) of candidate solutions. They do not require, for example, gradients or Hessian matrices, the computation of which might be prohibitively difficult or impossible in some problems.

\item \emph{Speed}. Apart from the intrinsically high speed with which EAs tend to explore large parameter spaces, they are embarrassingly parallel: that is, very little effort is required to transform a serial implementation into a parallel implementation. Thus they are well-suited to exploiting high-performance hardware such as multi-core workstations, graphics processing units (GPUs), clusters, etc.: see Section \ref{sec:GPUs}. To be sure, canonical EAs (GAs, ESs, etc.) do have a reputation for being inefficient local optimisers \citep{Charbonneau:1995,Michalewicz:1996}, but this problem can readily be mitigated either by coupling an EA to a dedicated local optimiser -- e.g.\ a gradient-based method -- or, as suggested in Section \ref{section:EMMA} below, by making simple tweaks to an EA's underlying mechanisms, so as to endow it with local optimisation capabilities.

\item \emph{Versatility}. A single EA-based optimiser can be expected to yield `good enough' results on a very wide class of problems -- from a problem as simple as fitting a three-parameter Gaussian to some data, to one as complex as choosing a molecular configuration to minimize a Buckingham potential with hundreds of parameters \citep{Wehrens:1998} -- and it is easy to incorporate problem-specific knowledge into an EA-based solver \citep{Haupt:2004}.
\end{enumerate}

The widespread adoption of EAs in fields such as engineering, chemistry, biology, and economics bears testimony to their many merits, and though their uptake in the physical sciences has been somewhat slower (at least partly because the theoretical understanding of their workings is still quite limited), recently they have already been used with great success in many branches of astronomy and astrophysics \citep{Rajpaul:2011}. In fact, a classical genetic algorithm -- PIKAIA -- has already been tested in the context of microlensing modelling \citep{Kubas:2005}, although it has not seen much use in recent years (at least, not in the microlensing community); unlike PIKAIA, though, the new algorithm presented here is tailored for high performance numerical optimisation, rather than for more general and pedagogical purposes.
\subsection{A new algorithm for fitting microlensing events}\label{section:EMMA}
The algorithm developed by the author for fitting microlensing lightcurves combines features of extant evolutionary algorithms (GSs, ESs) with some novel ones. The algorithm's main features are the following (a more detailed examination of the `nuts and bolts' of the algorithm is presented in Appendix A):
\begin{enumerate}
	\item a real-valued representation of solutions is used (i.e.\ parameter domains are not discretised, as would've been the case with e.g.\ a binary-coded GA);
	\item fitness-ranking can, in general, be performed according to arbitrary Bayesian priors and likelihood functions;
	\item individuals are chosen for reproduction via a variant of the well-known \emph{tournament selection} mechanism \citep[e.g.][]{Miller:1995};
	\item reproductive pairings comprise exactly two parent solutions, and give rise to exactly two offspring;
	\item the offspring are constructed in one of three ways -- as random interpolants of their parents, as exact duplicates of their parents, or as solutions where each parameter is drawn randomly but without modification from one of the two parents;
	\item both coarse-grained (`jump') and fine-grained (`creep') mutation operators are used, facilitating respectively global and local optimisation, with the mutation rates being dynamically adjusted in response to changes in population fitness; and
	\item the evolutionary sequence is restarted whenever the population stagnates around an optimum.
\end{enumerate}

The algorithm runs autonomously in the sense that, given a set of data and a general model (e.g.\ the SBLM) to be fit to the data, the algorithm will search intelligently\footnote{The algorithm's `intelligence' stems from its ability to exploit (via e.g.\ the reproduction and fine-grained mutation operators) structure  in the search space to guide its search, as well its ability to adapt its own control parameters in response to the progress of the search.} through the model parameter space to find solutions compatible with the data, without requiring any intervention from or assistance on the part of the user. 

The algorithm's development was informed by tests carried out not only on single- and binary-lens fitting problems, but also on a range of more general nonlinear optimisation problems \citep{More:1981, Haupt:2004}. The use of the dual mutation operators and dynamically-adjusted mutation rates means that, unlike more traditional EAs, the algorithm fares well at both global exploration and local optimisation. The jump operator allows for large jumps through the entire parameter space, whereas the creep operator is more conservative, and introduces only small (log-uniformly distributed) perturbations to existing solutions. The hybridised operators used for offspring construction are also more compatible with local fine-tuning than the canonical genetic crossover operator, which is well known to be antithetical to local optimisation \citep{Eiben:1998}. Finally, the algorithm was designed in an array programming paradigm (wherein all operators are applied to a single population matrix rather than to individual solutions) and, as such, the main algorithm is able to comprise fewer than $100$ lines of code when implemented in high-level language like \textsc{Matlab} or Python. 
\section{Fitting experiments and results}\label{section:experiments}
To test the EA, it was pitted against other optimisation algorithms and given the task of fitting several hundred simulated binary-lens lightcurves, each containing between one and four peaks (SBLM lightcurves with more than four peaks are theoretically possible though, assuming the parameter ranges in Table \ref{tab:params}, very rare.) The primary aim of the experiments was to study the feasibility of using the EA to explore binary model parameter spaces, and to quantify the algorithm's relative search efficiency/fitting success.
\subsection{Algorithms compared in tests}
The following four algorithms are compared in the tests that follow.
\begin{enumerate}
	\item \emph{Grid-search algorithm}. A simple brute-force approach, this algorithm evaluates systematically a `grid' of points, i.e.\ a predefined, uniformly-sampled subset of the model parameter space. The grid can be made arbitrarily fine (so that optimal solutions can be located with arbitrary accuracy) but computation time, which increases linearly with the total number of gridpoints, places a practical limit on the number of gridpoints that can be tested. In the fitting experiments, each dimension of the parameter space was discretised into the same number of linearly-spaced gridpoints, with the total number of gridpoints constrained by a predetermined limit on the fitting time. For example, if $10^{6}$ trial solutions could be evaluated within the time limit, and there were 7 parameters to be fit (as is the case for the SBLM), \textcolor{black}{$\ln{10^6}/\ln{7}\sim7$} different values of each parameter would've been tested.
	
	For reasons outlined in Section \ref{sec:difficulties}, such an algorithm should not be expected to be suitable for efficient modelling of microlensing events; still, it is included in these tests as a foil for the more sophisticated algorithms. Moreover, grid search algorithms \emph{have} found some (admittedly limited) use in the context of microlensing modelling \citep[e.g.][]{Kubas:2005,Bennett:2010,Gaudi:2010}, which underscores the paucity of search/optimisation algorithms, other than brute-force approaches, capable of reliably making headway on microlensing modelling problems.
	
	\item \emph{Iterated simplex algorithm}. The `downhill simplex method', also known as the `amoeba algorithm' or the `Nelder-Mead method', is an exceptionally popular nonlinear optimisation technique, due to \citet{Nelder:1965}. The downhill simplex method is used primarily for local optimisation, although it is endowed with some global optimisation capabilities, and like many other heuristic algorithms -- including evolutionary algorithms -- its operation is based purely on the \emph{evaluation} of a fitness/objective function; it does not, for example, depend on numerical or analytical gradients of such a function.

Given an optimisation function in $n$ free variables, the method starts by computing the values of the function at $n+1$ different points, which form the vertices of a general simplex. Then, by extrapolating the behavior of the function from the measurements made at each test point, the algorithm rapidly adapts the simplex to the local landscape, in such a way that the simplex `explores' its way downhill, until finally contracting onto a local minimum.

In the microlensing fitting experiments, the modern downhill simplex algorithm of \citet{Lagarias:1998} was used. As demonstrated by \citet{Charbonneau:2002b}, the use of multiple, short simplex runs (hereafter an `iterated simplex' approach), rather than one very long simplex run, greatly enhances performance on difficult \emph{global} optimisation problems: accordingly, in the microlensing fitting experiments, a number of random starting points were chosen in the search space, and at each point, a new downhill simplex optimiser was deployed. An individual simplex run was terminated when convergence to a local optimum was achieved, or when $10^4$ iterations had been reached. As with the grid search approach, the total number of starting points considered was limited only by the available computational time. 

Given the popularity, speed, and \emph{modus operandi} of the amoeba method, the iterated simplex approach constitutes a very reasonable competitor for an evolutionary algorithm.
	\item \emph{Evolutionary algorithm}. This is the newly-developed algorithm outlined in Section \ref{section:EMMA}. Default algorithm parameters, as discussed in Appendix A (viz.\ a fixed population size of $N_\textrm{pop}=1000$, a starting jump mutation probability of $p_\textrm{mut}=1\%$, etc.), were used in the fitting experiments.

Although the algorithm was designed to have both global exploration and local optimisation capabilities, its fits can very easily be fine-tuned by a \emph{dedicated} local optimiser such as the amoeba method described above. Therefore, the performance of the algorithm both with and without the benefit of amoeba fine-tuning was studied; in the former case, the amoeba method was used to fine-tune only one (viz.\ the fittest) solution returned by the evolutionary algorithm.
		\item \emph{Artificial neural network} (ANN). `Artificial Neural Networks' (ANNs, sometimes referred to simply as `neural networks') are mathematical/computational models that are inspired by biological neural networks. In an ANN, simple computational processing elements (`artificial neurons') are joined together to form a complex network of processing nodes, mimicking the structural and functional aspects of biological neural networks. ANNs are popular in the fields of regression, classification, pattern recognition, and decision-making; possibly the most attractive feature of ANNs, however, is their ability to be used as arbitrary function approximation mechanisms that can `learn' from observed data \citep{Michalewicz:2000}. In other words, they can perform autonomous, black-box modelling of the unknown (and possibly very complex) functional relationship between a set of input and output vectors, e.g.\ the relationship between a set of microlensing lightcurves, and their underlying model parameters.

The ANN considered in the fitting experiments that follow is a sophisticated algorithm, developed by \citet{Vermaak:2003,Vermaak:2007} specifically for the purpose of fitting SBLM lightcurves. Unlike the other algorithms discussed here, Vermaak's ANN-based fitter does not perform any iterative search through parameter space; instead, the mapping from lightcurves to model parameters is approximated directly by a complicated function derived from a very large set of training lightcurves. Although the ANN does require several hours to train, once-trained, fits can be generated very rapidly.

In many senses, the philosophies underpinning the ANN-based and EA-based fitting approaches are antithetical: the neural network is highly optimised for the fitting of SBLM lightcurves -- indeed, it is a truly bespoke algorithm -- but, unless it is retrained from scratch, it can \emph{only} fit SBLM lightcurves, and it certainly cannot handle more general optimisation problems. As with the the EA, however, the ANN's output can readily be fine-tuned by a dedicated local-optimiser; accordingly, the ANN's performance both with and without the benefit of amoeba optimisation is considered in the fitting experiments that follow.
		
\end{enumerate}

The fits performed by the grid search, the iterated simplex and the evolutionary algorithm were all guided by the minimisation of a $\chi^2$-statistic. It is well known that such a maximum-likelihood approach leads to biased parameter estimates \citep{Dominik:2008}; however, as already noted, the EA is fully-equipped for dealing with arbitrary Bayesian priors and likelihood functions. In fact, the assumptions of flat priors and Gaussian noise conveniently renders the more general fitting problem equivalent to that of a $\chi^2$-minimisation problem \citep{Press:2007}, though in principle the EA could quite easily be used to obtain e.g.\ maximum \emph{a-posteriori} parameter estimates instead. 

Although the ANN-based fitting did not involve the direct computation of any chi-square statistic (or indeed any goodness-of-fit statistic) during fitting, its training was based on a chi-square metric, so in the context of these fitting experiments, the ANN might be thought of as an indirect means of attempting a chi-square minimisation.

It should also be noted that none of the algorithms tested here explicitly incorporated any information specific to the microlensing modelling problem; this is discussed in more detail in Section \ref{sec:PSinfo}.
\subsection{Test setup}
Following \citet{Vermaak:2003}, the SBLM with parameter ranges as in Table \ref{tab:params} was used to generate lightcurves comprising $100$ magnification points, starting at a random time $t_{\rmn{start}}$, where
\begin{equation}
- 3 < \frac{{{t_{{\rmn{start}}}} - {t_m}}}
{{{t_E}}} <  - 2,
\end{equation}
and ending at a random time $t_{\rmn{end}}$, where
\begin{equation}
2 < \frac{{{t_{{\rmn{end}}}} - {t_m}}}
{{{t_E}}} < 3.
\end{equation}

The use of synthetic data (where input parameters are known exactly) meant that fitting success could be gauged not only in terms of goodness-of-fit, but more importantly, also in terms of adequate sampling of the `correct' regions of parameter space. Accordingly, a fitting sequence was deemed successful if it returned at least one model satisfying $\Delta\chi^2/\nu<1$, with errors in all fitted parameters of less than 10 percent; a weaker criterion of $\Delta\chi^2/\nu<1$ only, corresponding to a good fit(s), but not necessarily to closeness in parameter space, was also considered (here $\Delta\chi^2:=\chi^2-\chi^2_\rmn{true}$, where $\chi^2$ is the chi-square goodness of fit statistic for the lightcurve associated with a given solution/model, $\chi^2_\rmn{true}$ is the same statistic for the input lightcurve, and $\nu$ is the number of degrees of freedom for the fit).

The fitting experiments were carried out on a modern workstation with a six-core, 12-threaded, $3.6$~GHz processor, and 6~GB of 1333~MHz ECC RAM. To level the playing fields, all algorithms except for the ANN were parallelised, and restricted to $20$~minutes per fit (the ANN generated fits far more rapidly, and in any case, would not have benefited from a longer run time). 

In the case of the grid search algorithm, parallelisation entailed using all the available processor threads to evaluate twelve grid-points concurrently; in the case of the iterated simplex algorithm, twelve downhill-simplex runs were carried out in parallel. Parallelisation of the EA was equally trivial: each generation, twelve candidate solutions from the evolutionary population were evaluated concurrently. Since the computational expense of evaluating candidate solutions (mapping parameters to lightcurves and comparing the resultant lightcurves to observed/simulated data) dwarfed the expense of applying evolutionary operators (selection, reproduction, mutation, etc.) to the population, extremely efficient parallelisation was achieved by parallelising only one or two lines of the EA's code, i.e.\ those that called external, microlensing-related functions. The parallelisation of EAs is discussed in more detail in Section \ref{sec:parallel}.
\subsection{Results: noise-free lightcurves}
In the first test, lightcurves were uniformly-sampled and noise-free; again for consistency with the formalism and experiments of \citet{Vermaak:2003}, a value of $\sigma^{(k)}=0.01$ was adopted\footnote{Roughly speaking, this scaling -- though otherwise arbitrary -- meant that fits satisfying the $\Delta\chi^2/\nu<1$ criterion looked good `by eye'.} for the purposes of calculating $\Delta\chi^2/\nu$. Although this $\Delta\chi^2/\nu<1$ criterion has little statistical value in its own right, it does at least provide a means to quantify how closely a fitted lightcurve matches an input one (since, in the noise-free case, $\Delta\chi^2$ is proportional to the mean squared error for the fit), and thereby to rank the different algorithms.

Results from this test are presented in Table \ref{tab:results1}.

Sans the benefit of the amoeba-based local optimisation, the accuracy of both the grid search and the ANN-based fits was unacceptably poor (success rate close to $0\%$). The addition of the local optimisation improved the algorithms considerably, and when viewed in terms of success rates (as high as $70\%$ for the ANN) the results seem quite respectable; still, for critical work, a failure rate of around $30\%$ is far from ideal.

\begin{table*}
\begin{minipage}{116mm}
 \caption{Comparison of several techniques used to fit randomly-generated, noise-free binary lightcurves. A fitted model was deemed successful if it had $\Delta\chi^2/\nu<1$ \emph{and} errors in all fitted parameters $<10\%$; success rates using the weaker criterion of $\Delta\chi^2/\nu<1$ only are shown in parenthesis.}
\label{tab:results1}
\begin{tabular}{cccccc}
\hline
\multicolumn{ 1}{c}{Method} & \multicolumn{ 1}{c}{Reference} &    \multicolumn{ 4}{c}{Percentage successful fits} \\

\multicolumn{ 1}{c}{} & \multicolumn{ 1}{c}{} &     1 peak &    2 peaks &    3 peaks &    4 peaks \\
\hline
Grid search &  This work &      1 (1) &      0 (0) &      0 (0) &      0 (0) \\

Iterated simplex &  This work &    43 (98) &    36 (93) &    26 (78) &    16 (50) \\

Artificial neural network by itself & Vermaak &      4 (6) &      0 (0) &      0 (0) &      0 (0) \\
Artificial neural network with amoeba & Vermaak &    70 (96) &    70 (82) &    62 (72) &    68 (76) \\
Evolutionary algorithm by itself &  This work &   93 (100) &    96 (98) &    90 (93) &    89 (89) \\
Evolutionary algorithm with amoeba &  This work &   97 (100) &   98 (100) &    97 (98) &    94 (94) \\
\hline
\end{tabular}
\end{minipage}  
\end{table*}
\begin{table*}
 \begin{minipage}{116mm}
  \caption{Comparison of the EA with a conventional algorithm for fitting randomly-generated, noisy binary lightcurves, including random temporal gaps in the data, and blended light. Success criteria as in Table \ref{tab:results1}.}
  \label{tab:results2}
\begin{tabular}{cccccc}
\hline
\multicolumn{ 1}{c}{Method} & \multicolumn{ 1}{c}{Reference} &    \multicolumn{ 4}{c}{Percentage successful fits} \\

\multicolumn{ 1}{c}{} & \multicolumn{ 1}{c}{} &     1 peak &    2 peaks &    3 peaks &    4 peaks \\
\hline
Iterated simplex &  This work &  \color{black} 39 (95) &   \color{black}  29 (84) &    \color{black}13 (47) &    \color{black} 12 (42) \\

Evolutionary algorithm with amoeba &  This work &   \color{black} 95 (98) &  \color{black}   96 (99) &   \color{black} 90 (97) &  \color{black}  87 (99) \\
\hline
\end{tabular}  
\end{minipage}  
\end{table*}

Far more impressive was the EA which, coupled with the amoeba method, was easily the most accurate of all the algorithms: it failed to yield a very accurate fitted model for fewer than $4\%$ of the lightcurves. Even without the additional amoeba-based optimisation, the EA performed extremely well in its own right, which bears testimony to the EA's strength as both a global and a local optimiser. Additionally, in the handful of cases where the EA (with or without the amoeba method) failed, allowing the algorithm to run a little longer (e.g.\ for $30$~minutes, instead of the arbitrarily-imposed $20$-minute limit) always yielded successful fits. The same concession applied to the iterated simplex algorithm, but the time required to ensure successful fits was invariably much longer, i.e.\ on the order of several hours.

All algorithms fared better when fitting simpler (single- or double-peaked) lightcurves than when fitting the more complex ones, although for the EA this difference was very small. Indeed, it is quite surprising just how accurately the EA was able to home in on model parameters (in the case of successful fits, almost always to an accuracy of much better than $1\%$), even when the lightcurves had little apparent structure.
\subsection{Results: noisy, blended lightcurves\label{sec:noisy}}
In order to make the fitting problem more challenging, the following changes were made:
\begin{enumerate}
	\item temporal sampling of lightcurves was randomised by drawing observation times from a rectangular distribution on $(t_{\rmn{start}},t_{\rmn{end}})$, thus allowing for gaps, corresponding  e.g.\ to bad weather, in the data;
	\item Gaussian noise, as per the model defined by Eqn.\ \ref{eq:noiseModel}, was added to all datapoints;
	\item the possibility of significant blending was allowed for.
\end{enumerate}
Allowing for blending (wherein the presence of unlensed light, along the line-of-sight to the lens, dilutes the amplification signal of the source) meant extending the SBLM to include a new parameter, $f\in(0,1]$. If blended light is present, the observed amplification can, assuming constant blending, be computed as follows \citep{DiStefano:1995}:
\begin{equation}\label{eq:blend}
A(t) = f\cdot A_S(t) + (1-f),
\end{equation}
where $A_S(t)$ is the microlensing amplification of the source (the amplification that appears in Eqn.\ \ref{eq:A_s}), and the parameter $f$ is implicitly defined to be the ratio of the unlensed source flux, to the total unlensed flux (source flux, plus flux of all luminous sources along the line-of-sight to the lens, including flux from the lens itself) in the telescope beam. Although the blending effect is relatively trivial to calculate \citep[and in fact, a least-squares estimate for $f$ can be obtained directly from a simple linear equation; see e.g.][]{Jaroszynski:2001}, the presence of blended light does introduce ambiguities between competing models.

For consistency's sake, all other aspects of the experiment were left unchanged.

\begin{figure}
\begin{center}
\includegraphics[scale=0.8]{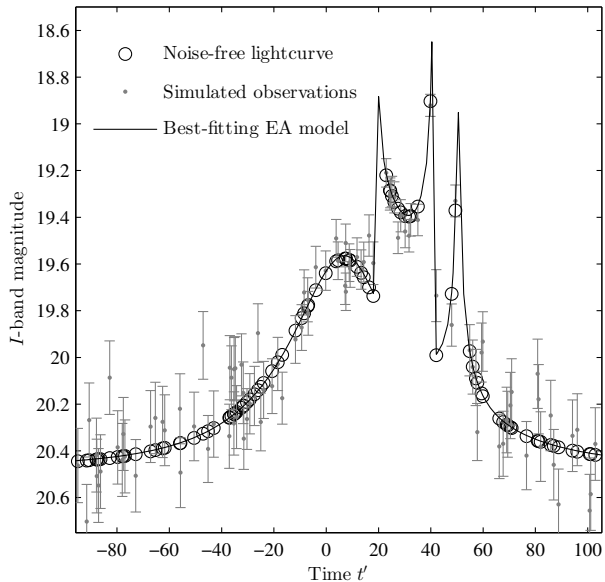}
\caption{A typical noise-free lightcurve generated via the SBLM, simulated observations of this lensing event, and the lightcurve for the model, as found by the EA, that best fits this (noisy) data.}
\label{fig:EAfit}
\end{center}
\end{figure}

Table \ref{tab:results2} gives an indication of how both the EA and the conventional algorithm fared on this somewhat more realistic problem (for brevity, this time only the performance of the amoeba fine-tuned algorithms is presented). The EA's performance remained excellent, with success rates generally in excess of $90\%$; this compares very favourably with the conventional algorithm's success rates of around $25\%$. Fig.\ \ref{fig:EAfit} gives an example of a typical fit performed by the EA.

Unfortunately no data were available to facilitate a direct comparison (i.e.\ using an identical noise model, randomised temporal sampling, and blending) with Vermaak's ANN; his conclusion, however, based on experiments with a very similar noise model was that noise has detrimental effects on the accuracy of ANN-based fitting, but that an ANN can at least be designed/(re)trained to mitigate these effects \citep{Vermaak:2007}. As such, even if one assumes that the ANN can perform as well with the noisy lightcurves as it did with the noise-free ones, the EA still emerges as the far more accurate algorithm.
\section{Discussion}
\subsection{Relative fitting performance of the EA}
Though the EA is, arguably, more computationally expensive than the ANN (once suitably trained, the ANN can perform a fit in under a minute; then again, actually training the ANN takes much longer, so it is not easy to compare the computational expense of the EA and the ANN on an equal footing), it certainly offers a search efficiency far beyond that of mere brute-force approaches, as evidenced by its clear outperformance of the grid search and iterated simplex approaches (both of which were calibrated to have the same computational footprint as the EA). The computational expense notwithstanding -- in any event, a fitting time of several minutes would be `fast enough' even for modelling ongoing events -- the EA offers a number of striking advantages over the ANN approach insofar as lightcurve analysis is concerned:
\begin{enumerate}
	\item the fitting accuracy of the EA is superior to that of the ANN;
	\item the algorithm itself is much simpler (it is relatively easy to code from scratch and to fine-tune an EA, which is certainly not the case with an ANN);
	\item the algorithm can be used straight `out of the box', and requires no training;
	\item the algorithm is far more robust since it doesn't need to be retrained/reconfigured to cope with noise, extensions to the basic lightcurve model, or even entirely different models (see Section \ref{section:future}); and
	\item the EA does not yield a single `all or nothing' fit - a single run yields very many fitted models, and the longer the algorithm is allowed to run, the better the chance of pinpointing the globally optimal models(s).
\end{enumerate}
\subsection{The importance of parallelisation\label{sec:GPUs}}

A fitting time on the order of several minutes sounds appealing, but unfortunately the simple binary-lens model adopted here neglects many effects often associated with real-world microlensing events, including parallax, instrumental, and finite-source effects. Dealing with some of these effects poses no significant difficulties, and requires only simple extensions to the basic SBLM; in particular, however, computing the magnification of an extended (as opposed to point-like) source can take around two orders of magnitude longer than the corresponding calculations under the point-source approximation \citep[e.g][]{Vermaak:2007,Gould:2008,Bozza:2010}. Unfortunately, finite-source effects generally \emph{cannot} be ignored when dealing with planetary signals \citep{Vermaak:2000}, so in such cases the claims about the algorithm's speed are seemingly invalidated. To make matters worse, the cadence of modern wide-field surveys can be nearly two orders of magnitude higher than in the lightcurves simulated here, \textcolor{black}{i.e.\ on the order of $15$~minutes \citep{Shvartzvald:2012}. The algorithm needn't be changed in any way to deal with higher cadences, though fitting times will, of course, increase in proportion to the increase in data rates.\footnote{Current work does, however, suggest the feasibility of using data compression to reduce full lightcurves to a smaller number of information-carrying datapoints, in order to accelerate analysis of binary events \citep{Heavens:2000,Hundertmark:2012}.}} This, combined with having to deal with finite-source effects, would suggest a total increase in fitting times by around four orders of magnitude!

Certainly, some performance gains can be expected from a meta-optimisation of the EA's control parameters (i.e.\ `fine-tuning' the algorithm specifically for fitting microlensing lightcurves); more importantly, however, EAs are highly amenable to parallel implementations, and they parallelise especially well when the cost of evaluating candidate solutions is high (as is certainly the case with microlensing modelling). Indeed, multi-core and multi-processor computers, clusters, grids, cloud computing platforms, and general-purpose GPUs (with upwards of a thousand streaming processors) are becoming increasingly prevalent in modern computing, and EAs are ideally suited to exploit such devices and platforms.

For example, by implementing an EA on a fairly high-end (though not top of the range) GPU card, \citet{Maitre:2009} managed to achieve a speed-up of roughly $100\times$ relative to the same EA running on a standard $3.6$~GHz processor, and advocated the use of multiple GPU cards to achieve even more dramatic speedups. A number of other authors have reported similar results \citep[see][and references contained therein]{Risco-Martin:2012}: \citet*{Pospichal:2010}, for example, managed to speed up a GA by a factor of nearly ten thousand, again using only fairly modest GPU hardware!

The possibility of such dramatic performance gains suggests that even when dealing with finite-source effects, it should be quite possible to constrain the EA's fitting times to under an hour. Porting existing code to GPU architectures does require quite specialised programming knowledge, but, given the potential gains, such an undertaking might well be a worthwhile investment of time and effort.

\subsection{Possible EA parallelisation schemes\label{sec:parallel}}
While EAs are not unique in being easy to parallelise (e.g.\ the grid search algorithm parallelises trivially!), EAs have the advantage of being amenable to a number of \emph{different} parallelisation schemes -- some straightforward, some more sophisticated -- and this provides scope for optimising an EA's performance for a specific hardware configuration and/or problem \citep{Haupt:2004}. 

The most trivial of all parallelisation schemes is to parallelise only the evaluation, each generation, of candidate solutions, with the rest of the algorithm -- which will usually have a very small computational footprint relative to the evaluation of candidate solutions -- left in serial form. Since selection occurs globally and within a single population, such EAs are usually referred to as `panmictic' or `micro-grained'. It was such an approach that was adopted for the EA discussed in this paper. (In problems where the computational cost of evaluating candidate solutions is relatively inexpensive, the cost of applying evolutionary operators cannot be ignored and efficient parallelisation would be harder to achieve.)

Another popular approach for parallelising EAs is to assign to each available processor an entire evolutionary population, and then to allow these populations to evolve more-or-less independently, perhaps allowing periodic `communication' (e.g.\ in the form of migration) between the populations. These so-called `coarse-grained' or `island' EAs are ideally suited for implementation on distributed memory MIMD (multiple instructions, multiple data) machines; in fact, even with all else being equal, island EAs tend to outperform single population EAs \citep{Gordon:1993}, with premature convergence being less of an issue, and optimal solutions usually being located more quickly. In this sense, parallelising an EA can lead not only to improved computational efficiency, but also to a more effective search algorithm!

A third major class of parallel EA is the `cellular' or `fine-grained' EA, which can be thought of as being intermediate to the two aforementioned classes. In such EAs, the evolutionary operators are decentralised, and each processor handles a number of very small evolutionary populations (perhaps containing only one or two individuals), each of which can interact with a number of `nearby' populations. Such an implementation is a often a natural choice on an SIMD (single instruction, multiple data) computer.
\subsection{Incorporation of problem-specific information\label{sec:PSinfo}}

As already noted, none of the algorithms tested incorporated any information specific to the structure of the microlensing modelling problem, e.g.\ knowledge of lightcurve morphologies or of the topologies of underlying critical curves and source trajectories. Incorporating such prior knowledge could be used to make most of the algorithms, except the ANN, more efficient: that is, rather than letting them search `blindly' through the full SBLM parameter space, problem-specific information could be used to reduce the volume of the space of feasible parameters, and could provide clues as to where good solutions are likely to be found \citep{Kains:2009}. Insofar as this work goes, the main reason microlensing-specific information was not incorporated was to facilitate a more fair comparison of the other algorithms with Vermaak's ANN, into which the incorporation of problem-specific information would be very difficult. Still, some more general arguments could be made in favour of the non-inclusion of such information. Incorporating knowledge of the SBLM and its associated lightcurves, for example, would increase the complexity and reduce the generality of the algorithms, and different information would need to be incorporated if one wanted to fit, say, a triple-lens model instead of a binary model. Indeed, if the computational power is available for an algorithm to perform fits in a reasonable amount of time, the manual incorporation of problem-specific information would be, if nothing else, an unnecessary expense of human effort.\footnote{By way of analogy, knowledge of the properties of polynomials can often be used to find some of the roots of certain classes of polynomial but, especially when dealing with higher-order polynomials, it is usually far more straightforward and efficient to find the roots with a general root-finding algorithm than to try to apply analytical results \citep{King:1996,Press:2007}.}

Indeed, we are at a stage where the computational power is available to drive intelligent search/optimisation algorithms that can rival the active modelling efforts of humans (using analytical techniques, problem-specific information, etc.), even on very difficult problems. This consideration might be a largely academic one in the context of modelling binary events \citep[where bespoke algorithms that incorporate problem-specific information have already been developed and implemented with success, e.g.][]{Kains:2009}, but this consideration would apply to any difficult modelling problem, including ones where bespoke algorithms have not yet been developed.

\subsection{EA vs.\ Monte Carlo optimisation}

It is worth mentioning that Monte Carlo methods are ubiquitous in the field of microlensing modelling \citep{Sumi:2010,Shin:2011,Skowron:2011}. During the course of this work, a Markov chain Monte Carlo (MCMC) method, incorporating an adaptive Metropolis sampler and a delayed rejection mechanism \citep{Haario:2006}, was deployed and used to perform fits on single-lens events. The MCMC method was found to be nearly two orders of magnitude slower than the EA, and so no in-depth tests of the MCMC method were carried out for binary-lens events. In any event, even if an EA can perform initial fits more quickly, Monte Carlo would remain indispensable e.g.\ for estimating uncertainties in fitted parameters found by the EA \citep{Press:2007}.

\subsection[Future work]{Future work} \label{section:future}
Although some of the merits of EAs have been extolled in this paper, this work serves merely as a proof of concept for autonomous, EA-based microlensing modelling: the EA's relative performance is promising, but it remains to be seen how it will fare when faced with more complex microlensing modelling problems. Indeed, the simple binary-lens model and simulation parameters adopted here \citep[primarily for the sake of allowing a direct comparison with the results of][]{Vermaak:2003,Vermaak:2007} led to the exclusion of many effects usually associated with real-world microlensing events, including finite-source, parallax, and lens orbital-motion effects. Some of the parameter ranges used here would need to be extended -- or, better yet, replaced with more realistic Bayesian priors that avoid hard limits -- to cover cases where the secondary lens is an exoplanet. Finally, to be more representative of modern, high-cadence surveys, temporal sampling of simulated lightcurves would need to be increased by about two orders of magnitude \citep{Shvartzvald:2012} -- this would increase fitting times, but should also help to resolve some model ambiguities.

Still, since the EA managed to move effortlessly from single-lens to binary-lens to noisy binary fitting problems, one can speculate that fitting problems involving an extended SBLM should not pose major obstacles to the EA. Of course, given a more complex modelling problem, one would need to sacrifice some fitting accuracy, and/or harness more computing power (as suggested in Section \ref{sec:GPUs}) if one wanted to leave fitting times unchanged.

An immediate extension to the present work will be an investigation by the author of the EA's utility in modelling ongoing microlensing events, the hope being that it might provide a fast and accurate means of \emph{forecasting} important features in real-world lightcurves. This in turn could facilitate better-informed observations, and ultimately, more useful data. In fact, preliminary results from this work (using more realistic simulated lightcurves, featuring greatly expanded parameter ranges, much higher temporal sampling, blended light, etc.) have been promising -- the EA does seem well-suited to the efficient location of solutions compatible with existing data, and once such solutions have been found, it is a straightforward task to generate predictions of e.g.\ the time of a caustic crossing. Successful EA-based forecasts have been already been obtained -- allowing fitting times on the order of an hour or two, \textcolor{black}{though with the general algorithm otherwise set up in much the same way as described in this work} -- on a small selection of real events (using truncated versions of their completed lightcurves), including MOA-2011-BLG-197/OGLE-2011-BLG-0265,\footnote{See http://www.stelab.nagoya-u.ac.jp/\%7Esumi/anomaly/2011/.} lending some credence to the general feasibility of an EA-based fitting approach. However, it remains to quantify rigorously the extent to which intrinsic model degeneracy and nonlinearity hinder reliable forecasts, both with and without the incorporation into the forecasting of problem-specific information into the modelling. Results from this work are to appear in a dissertation (currently in prep.) by the author.
\section{Conclusions}
A novel evolutionary algorithm was introduced and was shown to be well suited to fitting binary-lens microlensing lightcurves. Despite the difficulty of this fitting problem, the algorithm yielded excellent fitting accuracy whilst maintaining a relatively modest computational footprint, and offering a number of other desirable properties (robustness, versatility, trivial parallelisation, etc.). As such, this work provided proof of concept for the use of an evolutionary algorithm as the basis for real-time, autonomous modelling of microlensing events. It was noted, however, that further work is required to investigate how the algorithm will fare when faced with more complex and realistic microlensing modelling problems.

Indeed, this work could be extended in a number of ways, in particular by relaxing some of the assumptions made here, and also by investigating how an evolutionary algorithm will fare at trying to predict features in the lightcurves of \emph{ongoing} microlensing events. Early investigations in this regard have been promising.

Though evolutionary algorithms can by no means be the `final word' in microlensing modelling, a simple evolutionary algorithm such as the one presented here could serve as a useful addition to the toolbox of a microlensing modeller, or indeed, of any astronomer working with difficult model-fitting problems.
\section*{Acknowledgments}
This work was supported financially by the University of Cape Town and the National Research Foundation. The author wishes to thank the anonymous referee for his/her helpful comments, constructive criticism, and patience; similarly, the author expresses gratitude to Dr John Menzies and Dr Ian Stewart.
\bibliographystyle{mn2e} 																	
\bibliography{mnPaper}
\label{lastpage}
\appendix
\section{Mechanics of the new algorithm}
This appendix sketches the `nuts and bolts' of the new evolutionary algorithm introduced in Section \ref{section:EMMA}, albeit without providing a detailed rationale for its various features.\footnote{Original \textsc{Matlab} code for the algorithm is available from the author; the code itself is not presented here because, though very compact, it would make little sense without detailed documentation to explain the various external functions called by the code, \textsc{Matlab} syntax and operators, etc. In any event, the description of the algorithm given here is sufficient to construct, in any language, a working version of the algorithm.}  An in-depth discussion/analysis of the algorithm is to be included in a dissertation by the author (currently in prep.); suffice it to say, however, that most aspects of the algorithm's design were informed by extensive empirical testing, as well as theoretical considerations.

The central dynamical quantity used by the algorithm, as with most EAs, is an (encoded) \emph{population matrix}, $\mathbb{P}$:
\begin{equation}
\mathbb{P} = \mathbb{P}(t) = {[p_{i,j}]}_{N_\textrm{pop}\times N_\textrm{par}}.
\end{equation}
$\mathbb{P}(t)$ is constructed so that $p_{i,j}\in[0,1]$ \emph{encodes} the value of the $j^\textrm{th}$ physical parameter associated with the $i^\textrm{th}$ individual (trial solution) in the $N_\textrm{pop}$-member evolutionary population, denoted $a_{i,j}$, at generation $t\in\{1,2,\ldots,N_\textrm{gen}\}$. By default, a population size of $N_\textrm{pop}=1000$ is assumed; the number of fitting parameters, $N_\textrm{par}$, will be fixed by the problem at hand (e.g.\ $N_\textrm{par}=8$ for the fitting experiments in Section \ref{sec:noisy}).

$\mathbb{P}$ may be thought of as representing an encoded ensemble of trial solutions, with a row vector $p_{i,*}$ representing a single trial solution, and an individual matrix element ($p_{i,j}$) representing a small `chunk' or component of a trial solution. The relationship between $\mathbb{P}(t)$ and its decoded/physical counterpart, denoted $\mathbb{A}(t)$, can be represented thus:
\begin{equation}
[{p_{i,j}}] = \mathbb{P}\begin{array}{*{20}{c}}
   {\xrightarrow{{{\text{decoding}}}}}  \\
   {\xleftarrow[{{\text{decoding}}}]{{}}}  \\
 \end{array} \mathbb{A} = [{a_{i,j}}];
\end{equation}
the actual encoding/decoding process is explained in Section \ref{section:enc}.

The basic evolutionary sequence used by the EA is as follows:
\begin{enumerate}
	\item Initialise $\mathbb{P}(t=1)$.
	\item Iterate $t=1,2,\ldots,N_\textrm{gen}$; for each $t$:
		\begin{enumerate}
			\item decode $\mathbb{P}(t)\to\mathbb{A}(t)$, and pass $\mathbb{A}(t)$ to an external fitness function;
			\item rank $\mathbb{P}(t)$ based on the fitness returned by the external function;
			\item select solutions for reproduction;
			\item produce offspring and insert them into $\mathbb{P}(t)$;
			\item apply jump and creep mutations to $\mathbb{P}(t)$, and
			\item check whether the evolution has stagnated, and if so, make appropriate adjustments to the algorithm.
		\end{enumerate}
	\item Collate and output results, as required.
\end{enumerate}
The total number of evolutionary generations, $N_\textrm{gen}$, can be used to control how long the algorithm runs: larger values of $N_\textrm{gen}$ will lead to longer run-times, though with more solutions being evaluated. (Alternatively, the algorithm can be allowed to run for an arbitrary number of generations, until some user-defined convergence criterion is achieved.)

The algorithm's basic working scheme is conceptually straightforward, and most of its features overlap broadly with those of the canonical genetic algorithm; many of its finer details and features -- some unique to the algorithm, some inspired by other EAs -- are, however, non-canonical. 

Each step in the aforementioned working scheme is described in the subsections below.

\subsection{Initialisation}
Initialising the population matrix entails simply assigning random variates with a standard uniform distribution, $\mathcal{U}(0,1)$, to half of the elements in $\mathbb{P}$. The remaining elements are then assigned the complements of the values already assigned. That is to say, if one element in the population matrix is assigned the value $X\in[0,1]$, another random element will be assigned the value $(1-X)\in[0,1]$.

The initialisation scheme ensures a uniform and unbiased initial sampling of the entire (encoded) parameter space. (It is possible to change the initialisation scheme to incorporate prior information via a bias in the initial population, although in practice, a more robust and theoretically-sound approach is to incorporate prior information into the external fitness function.) 
\subsection{Encoding/decoding}\label{section:enc}
The following bicontinuous linear transformation is used to relate a physical parameter value, $a_{i,j}$, to its encoded counterpart, $p_{i,j}\in[0,1]$:
\begin{equation}
\label{eq:EMMA-encoding}
{a_{i,j}} = {\alpha _j} + ({\beta _j} - {\alpha _j}){p_{i,j}},
\end{equation}
where it assumed that the physical parameter $a_j$ can be restricted to some domain ${\Omega _j} = [{\alpha _j},{\beta _j}]$, with $\alpha_j$ and $\beta_j$ known. For example, if $a_j$ is an angle, a typical domain might be ${\Omega _j}=[0,2\pi)$. Even if $a_j$ is in principle unbounded, physical considerations and/or prior knowledge associated with the problem at hand should always allow bounds to be placed on the parameter. For example, if $a_j$ represents a binary-lens mass ratio, a safe choice might be ${\Omega _j}=[10^{-6},1]$, say.

The interpretation of the encoded value $p_{i,j}$, as implicitly defined in Eqn.\ \ref{eq:EMMA-encoding}, is straightforward: $p_{i,j}$ represents the fractional position of $a_{i,j}$ along its `physical' domain $[{\alpha _j},{\beta _j}]$. It should also be clear that the algorithm works with real-valued representations of parameters ($p_{i,j}\in\mathbb{R}$), rather than discretising the parameter domains in any way, as would've been the case with e.g.\ a binary-coded genetic algorithm. 

The encoding scheme amounts to working with all parameters normalised to the \emph{unit} interval. The normalisation is largely a matter of convenience, and means that all parameters can be treated on an `equal footing' when applying evolutionary operators (without normalisation, a different mutation operator would be required for each different parameter, to ensure mutations don't produce parameter values outside the allowable domains; with normalisation, a single mutation operator can be applied to $\mathbb{P}$ as whole, making for simpler code). As for the normalisation specifically to the unit interval: the unit interval corresponds identically to the support of the standard uniform distribution -- which serves as a basis for sampling from most other probability distributions -- a fact that simplifies the coding of the stochastic aspects of the algorithm. Indeed, it is possible to forego parameter encoding altogether, though this would come at the cost of more complex code for handling mutation operators, crossover operators, etc.

\subsection{Fitness evaluation and ranking}
Fitness evaluation entails passing the decoded trial solutions in $\mathbb{A}(t)$ to some external, problem-specific function that assigns a well-defined figure of merit, or fitness, to each solution. In the case of data modelling, this figure of merit will usually be related to a normalised fitting error -- e.g. a $\chi^2$ statistic (more specifically, something like $1-\chi^2$ or $1/\chi^2$ so that, as per convention, better solutions can be associated with a higher fitness). An important point is that fitness-ranking can actually be performed according to arbitrary Bayesian priors and likelihood functions, say -- in fact, the algorithm itself makes absolutely no assumptions about the statistical paradigm (e.g.\ Bayesian vs.\ frequentist) in which data modelling is taking place.

Once the solutions have been evaluated, the rows of the encoded population matrix $\mathbb{P}(t)$ are sorted so that the fittest solution occupies row 1, the next fittest solution row 2, and so on, with the worst solution in row $N_\textrm{pop}$. $\mathbb{A}(t)$ is not sorted as it is not be used again during generation $t$, and is simply recomputed during generation $t+1$. 
\subsection{Selection}
Once the quality of each trial solution has been ascertained, this information is used to select the solutions that will reproduce, i.e.\ the solutions that will have their `genetic material' (encoded chunks of solutions) propagated to the next generation. 

Reproductive pairings are picked via a variant of the well-known `tournament selection' mechanism \citep{Miller:1995}. The scheme used to choose any parent is as follows: \emph{draw a random sample of} ${N_{{\text{tourn}}}}\in \mathbb{N}^*$ \emph{values from the set} $\{ 1,2, \ldots ,{N_{{\text{pop}}}}\}$. \emph{Find the minimum value in that sample, and let this number be the \emph{rank} of a parent chosen for reproduction}. This scheme is repeated as many times as is necessary to produce the desired number of reproductive pairings. By default, a value of $N_{{\text{tourn}}}=\lceil N_{{\text{pop}}}/25 \rceil $ is used.

The basic idea behind tournament-style selection is that it corresponds to running a number of small tournaments or `fights' between solutions, with the `victor' of an individual tournament, viz.\ the fittest (best-ranked) solution in that particular tournament, being allowed to reproduce. The larger the tournament, the higher the \emph{selection pressure}, i.e.\ the less likely that poor solutions will be chosen for reproduction.

Note that, because the selection scheme is rank-based rather than fitness-based, the exact choice and normalisation of the fitness function (e.g.\ $1-\chi^2$ vs.\ $1/\chi^2$ vs.\ $1-\sqrt{\chi^2}$) is largely irrelevant to the algorithm, so that performance will generally not hinge on a `clever' choice of a fitness function.  More precisely, the algorithm's performance will be identical when using any two fitness functions that can be related via a monotone (order-preserving) transformation. 

\subsection{Reproduction}
Offspring are produced by combining the genetic material of parent solutions. For simplicity it is assumed that exactly two parents are involved in any reproductive pairing, and that each such pairing gives rise to two offspring. Assuming the parent solutions chosen for a particular pairing are $p_{i,*}$ and $p_{j,*}$, the algorithm will produce two offspring via the following two convex combinations of parent solutions:
\begin{equation}\label{eq:rep1}
{p_{i,*}} \cdot \mathbb{F} + {p_{j,*}} \cdot (\mathbb{I} - \mathbb{F}),
\end{equation}
and
\begin{equation}\label{eq:rep2}
{p_{j,*}} \cdot \mathbb{F} + {p_{i,*}} \cdot (\mathbb{I} - \mathbb{F}),
\end{equation}
where $\mathbb{F}=[f_{i,j}]$ is an $N_\textrm{par}\times N_\textrm{par}$ diagonal `weighting' matrix, with all elements constrained to the unit interval, and $\mathbb{I}$ is the $N_\textrm{par}\times N_\textrm{par}$ identity matrix. 

The actual choice of $\mathbb{F}$ determines the characteristics of the reproduction. For example, if $f_{i,i}=\tfrac{1}{2}$~$\forall \; i$, the offspring will simply be arithmetic averages of the parent solutions, as may easily be confirmed via Eqns.\ \ref{eq:rep1} and \ref{eq:rep2}; on the other hand, if each $f_{i,i}$ is assigned randomly, then the offspring will be correspondingly random interpolants of the parents solutions.

The algorithm assigns $\mathbb{F}$ as follows. \emph{For each reproductive pairing, with equal probability, assign diagonal elements of the weighting matrix $\mathbb{F}$ according to any one of the following three distributions}:
\begin{equation}
\left\{ \begin{array}{l}
 {f_{i,i}}\sim{\mathcal U}(0,1),\;{\rm{or}} \\
  \; \\ 
 {f_{i,i}}\sim{\textrm{Bin}}(1,\tfrac{1}{2}),{\rm{ or}} \\
 \; \\ 
 {f_{i,i}} = 1{\rm{ }}\; \forall \:i. \\ 
 \end{array} \right.
\end{equation}
The distribution used to make the assignment is chosen randomly for each reproductive pairing. In the first case, the genes of the offspring will be random \emph{interpolants} of the parental genes; in the second case the offspring's genes will be randomly drawn from the combined gene-pool of the parents, but no interpolation will occur; and the final case corresponds to asexual reproduction wherein offspring are identical copies of their parents (which is useful for giving the upcoming mutation operators multiple opportunities to fine-tune a solution that is already very good). Both the second and third schemes can be thought of as limiting cases of the first, more general scheme.

All solutions in $\mathbb{P}(t)$, save for the fittest one, are replaced by the newly-created offspring. This non-replacement of the fittest individual in $\mathbb{P}(t)$ is referred to as `elitism' in the literature \citep[e.g.][]{Michalewicz:2000}, and is a standard feature of most EAs. A practical implication of elitism is that the fitness of the fittest individual in the population will increase monotonically with $t$ (unless of course the evolutionary sequence is restarted, as discussed in Section \ref{sec:EMMA-stag} -- even in this case, however, the evolution history of the population will be saved, so in a global sense, the fitness of the best solution known will still increase monotonically with $t$).
\subsection{Creep and jump mutation}\label{sec:EMMA-mut}
All newly-created offspring are subject to fine-grained mutations, also known as `creep mutations'. These mutations randomly increase or decrease an encoded value by a randomly-chosen step with a log-uniform distribution:
\begin{equation}
{p_{i,j}} \to p_{i,j}^ + {\text{ or }}p_{i,j}^ - ,{\text{   }}\left\{ {\begin{array}{*{20}{c}}
   {p_{i,j}^ -  = {p_{i,j}} - ({p_{i,j}} - 0) \cdot X,}  \\
   {p_{i,j}^ +  = {p_{i,j}} + (1 - {p_{i,j}}) \cdot X}  \\
 \end{array} } \right\}
\end{equation}
where the actual assignment ($p_{i,j}^ +$ or $p_{i,j}^ -$) is chosen with equal probability, $\ln X \sim \mathcal{U}\left( {\ln \varepsilon ,\ln 1} \right)$, and $\varepsilon$ is the machine epsilon. Note that the creep mutations are fully compatible with the algorithm's parameter encoding scheme because $\forall\;X,p_{i,j}^ \pm\in[0,1]$.

In addition to the creep mutations, the offspring are exposed to the possibility of coarse-grained or `jump' mutations. With a jump mutation, an element of the population matrix is `flipped' to a completely random value on the unit interval:
\begin{equation}
{p_{i,j}} \to X \sim U(0,1).
\end{equation}
A jump mutation is assumed to happen with some small probability $0<p_\textrm{mut}\ll 1$, so that in a given generation, on average $p_\textrm{mut}\times N_\textrm{pop}\times N_\textrm{par}$ elements of $\mathbb{P}(t)$ are subject to a jump mutation. By default, $p_\textrm{mut}=1\%$ is used. (The mutation operators were designed to strike a balance between adequate global exploration and efficient local optimisation, and their design was informed both by very extensive empirical testing as well as theoretical considerations.) 
\subsection{Stagnation checks}\label{sec:EMMA-stag}
If the evolution is deemed to have stagnated -- as determined by any criterion, the default one being the fitness of the fittest individual in $\mathbb{P}(t)$ not having improved by more than $1\%$ over the past 10 generations -- the jump-mutation probability will be boosted (increased by $50\%$); conversely, if the evolution has \emph{not} stagnated, the jump-mutation probability will be throttled back (decreased by $50\%$). This autonomous boosting/throttling back of mutation probabilities, in response to the rate of improvement of solution fitness, serves to shift emphasis between global exploration (as mediated by the jump mutation operators) and local optimisation (as mediated, in part at least, by the more conservative creep mutation operators).

As a further step, if repeated boosts to the jump-mutation probability do not have the desired effect, i.e.\ stagnation flags are raised repeatedly, $\mathbb{P}(t)$ will be saved to memory, and the population will be reinitialised as was done at $t=1$, so that $\mathbb{P}(t+1)$ is a fresh evolutionary population, with no dependence on $\mathbb{P}(t)$.
\subsection{Outputs}
By default, the algorithm outputs all dynamical quantities (population matrices, associated fitness vectors, mutation rates, etc.) computed for all $t$. The outputs from the algorithm can, however, be tailored, with minimal effort, to suit one's needs: in simple problems, one might only be interested in a single `best solution' (and the quality of that solution), whereas with complex, multimodal problems, one might wish to obtain $\mathbb{A}(t)$ for all $t$, and the associated fitness vectors, e.g.\ $\chi^2(\mathbb{A})$, in order to construct confidence intervals for model parameters, say.
\end{document}